\documentclass[10pt,a4paper]{article}
\usepackage{amsmath}
\usepackage{amssymb}
\usepackage{amsfonts}
\usepackage{amstext}
\usepackage{amsbsy}
\usepackage[mathscr]{eucal}
\usepackage{graphicx}
\usepackage{color}
\newcommand{\beq}{\begin{equation}}
\newcommand{\eeq}{\end{equation}}
\newcommand{\beqa}{\begin{eqnarray}}
\newcommand{\eeqa}{\end{eqnarray}}
\newcommand{\ket}[1]{| #1 \rangle}
\newcommand{\bra}[1]{\langle #1 |}

\title{\Large\textbf{Concurrence classes for general pure multipartite states}}

\author{\textit{Hoshang Heydari}\\{\small\emph{Institute of Quantum
Science, Nihon University,}}\\ \small{\emph{1-8 Kanda-Surugadai,
Chiyoda-ku, Tokyo 101- 8308, Japan}}}
\begin{document}

\maketitle

\begin{abstract}
We propose concurrence classes for general pure multipartite
states based on an orthogonal complement of a positive operator
valued measure on quantum phase. In particular, we construct
$W^{m}$ class, $GHZ^{m}$, and $GHZ^{m-1}$ class concurrences for
general pure $m$-partite states. We give explicit expressions for
$W^{3}$ and $GHZ^{3}$ class concurrences for general pure
three-partite states and for $W^{4}$, $GHZ^{4}$, and $GHZ^{3}$
class concurrences for general pure four-partite states.
\end{abstract}

\section{Introduction}
Entanglement is an interesting feature of quantum theory, which in
recent years has attracted many researchers to quantify, classify
and to investigate its useful properties. Entanglement has already
some applications such as quantum teleportation and quantum key
distribution, and there will be new applications for this
fascinating quantum phenomenon. For example, multipartite
entanglement has the capacity to offer new unimaginable
applications in emerging fields of quantum information and quantum
computation. One of the widely used measures of entanglement for
 a pair of qubits is  the concurrence, which is directly related to
the entanglement of formation
\cite{Bennett96,Wootters98,Wootters00}. In recent years there have
been some proposals to generalize this measure to a general
bipartite state. For example, Uhlmann \cite{Uhlmann00} has
generalized the concept of concurrence by considering arbitrary
conjugation. Later Audenaert \emph{et al.}\cite{Audenaert}
generalized this formula in spirit of Uhlmann's work, by defining
a concurrence vector for a pure bipartite state. Another
generalization of concurrence was suggested by Rungta \emph{et
al.}\cite{Rungta01} based on a super operator called universal
state inversion. Moreover, Gerjuoy \cite{Gerjuoy} and Albeverio
and Fei \cite{Albeverio} gave an explicit expression of the
concurrence in terms of the coefficients of a general pure
bipartite state. It would therefore be interesting to be able to
generalized this measure from bipartite to a general multipartite
state, see Ref. \cite{Bhaktavatsala,Akhtarshenas,Wang}.
Quantifying entanglement of multipartite states has been discussed
in
\cite{Lewen00,Vedral97,Werner89,Plenio00,Hor00,Acin01,Bennett96a,Dur99,
ECKERT02,Dur00,Eisert01,Verst03,Pan}. In \cite{Hosh3,Hosh4}, we
have proposed a degree of entanglement for a general pure
multipartite state based on a positive operator valued measure (
POVM) on quantum phase. Recently, we have also defined concurrence
classes for  multi-qubit mixed states \cite{Hosh5} based on an
orthogonal complement of a POVM on quantum phase. In this paper,
we will construct different concurrence classes for general pure
multipartite states. Our concurrence classes vanish on the product
state by construction. For multi-qubit states, the $W^{m}$ class
concurrences are invariant under stochastic local quantum
operation and classical communication(SLOCC) \cite{Dur00}, since
orthogonal complement of our POVM are invariant under the action
of the special linear group. Furthermore, all homogeneous positive
functions of pure states that are invariant under determinant-one
SLOCC operations are entanglement monotones \cite{Verst03}.
However, invariance under SLOCC  for the $W^{m}$ class concurrence
for general multipartite states need deeper investigation. It is
worth mentioning  that Uhlmann \cite{Uhlmann00} has shown that
entanglement monotones for concurrence are related to antilinear
operators. The $GHZ^{m}$ class concurrences for multipartite
states introduced in this paper are not entanglement monotones
except under additional conditions. Thus, the $GHZ^{m}$ class
concurrences need further investigation. Classification of
multipartite states has been discussed in
\cite{Verst,Oster,Miyake,Miyake04,Wang}. For example, F.
Verstraete \emph{et al}. \cite{Verst} have considered a single
copy of a pure four-partite state of qubits and investigated its
behavior under SLOCC, which gave a classification of all different
classes of pure states of four qubits. They have also shown that
there exist nine families of states corresponding to nine
different ways of entangling four qubits. A. Osterloh and J.
Siewert \cite{Oster} have
 constructed entanglement measures for pure
states of multipartite qubit systems. The key element of their
approach is an antilinear operator that they called comb. For
qubits, the combs are invariant under the action of the special
linear group. They have also discussed inequivalent types of
genuine four-qubit entanglement, and found three types of
entanglement for these states. This result coincides with our
classification, where in section \ref{Sec4} we construct three
types of concurrence classes for four-qubit states.  A. Miyake
\cite{Miyake}, has also discussed classification of multipartite
states in entanglement classes based on the  determinant. He shown
that two states belong to the same class if they are
interconvertible under SLOCC. Moreover, the only paper that
addressed the classification of higher-dimensional multipartite
states is the paper by A. Miyake and F. Verstraete
\cite{Miyake04}, where they have classified multipartite entangled
states in the $ 2\times2 \times n$ quantum systems for (n $\geq$
4). They have shown that there exist nine essentially different
classes of states, and they give rise to a five-graded partially
ordered structure, including GHZ class and W class of 3 qubits.
Finally, A. M. Wang \cite{Wang} has proposed two classes of the
generalized concurrence vectors of the multipartite systems
consisting of qubits. Our classification is similar to Wang's
classification of multipartite state. However, the advantage of
our method is that our POVM can distinguish these concurrence
classes without prior information about in equivalence of these
classes under local quantum operation and classical communication
(LOCC).

Let us denote a general, multipartite quantum system with $m$
subsystems by
$\mathcal{Q}=\mathcal{Q}_{m}(N_{1},N_{2},\ldots,N_{m})$ $
=\mathcal{Q}_{1}\mathcal{Q}_{2}\cdots\mathcal{Q}_{m}$, consisting
of a state
\begin{equation}\label{Mstate}
\ket{\Psi}=\sum^{N_{1}}_{k_{1}=1}\sum^{N_{2}}_{k_{2}=1}\cdots\sum^{N_{m}}_{k_{m}=1}
\alpha_{k_{1},k_{2},\ldots,k_{m}} \ket{k_{1},k_{2},\ldots,k_{m}}
\end{equation}
 and, let
$\rho_{\mathcal{Q}}=\sum^{\mathrm{N}}_{n=1}p_{n}\ket{\Psi_{n}}\bra{\Psi_{n}}$,
for all $0\leq p_{n}\leq 1$ and $\sum^{\mathrm{N}}_{n=1}p_{n}=1$,
denote a density operator acting on the Hilbert space $
\mathcal{H}_{\mathcal{Q}}=\mathcal{H}_{\mathcal{Q}_{1}}\otimes
\mathcal{H}_{\mathcal{Q}_{2}}\otimes\cdots\otimes\mathcal{H}_{\mathcal{Q}_{m}},
$ where the dimension of the $j$th Hilbert space is given  by
$N_{j}=\dim(\mathcal{H}_{\mathcal{Q}_{j}})$. Moreover, let us
introduce a complex conjugation operator $\mathcal{C}_{m}$ that
acts on a general state $\ket{\Psi}$ of a multipartite state as
\begin{equation}
\mathcal{C}_{m}\ket{\Psi}=\sum^{N_{1}}_{k_{1}=1}\sum^{N_{2}}_{k_{2}=1}\cdots\sum^{N_{m}}_{k_{m}=1}
\alpha^{*}_{k_{1},k_{2},\ldots,k_{m}}
\ket{k_{1},k_{2},\ldots,k_{m}}.
\end{equation}
We are going to use this notation throughout this paper, i.e., we
denote a mixed pair of qubits by $\mathcal{Q}_{2}(2,2)$. The
density operator $\rho_{\mathcal{Q}}$ is said to be fully
separable, which we will denote by $\rho^{sep}_{\mathcal{Q}}$,
with respect to the Hilbert space decomposition, if it can  be
written as $ \rho^{sep}_{\mathcal{Q}}=\sum^\mathrm{N}_{n=1}p_{n}
\bigotimes^m_{j=1}\rho^{n}_{\mathcal{Q}_{j}}$,
$\sum^\mathrm{N}_{n=1}p_{n}=1 $,
 for some positive integer $\mathrm{N}$, where $p_{n}$ are positive real
numbers and $\rho^{n}_{\mathcal{Q}_{j}}$ denote a density operator
on Hilbert space $\mathcal{H}_{\mathcal{Q}_{j}}$. If
$\rho^{p}_{\mathcal{Q}}$ represents a pure state, then the quantum
system is fully separable if $\rho^{p}_{\mathcal{Q}}$ can be
written as
$\rho^{sep}_{\mathcal{Q}}=\bigotimes^m_{j=1}\rho_{\mathcal{Q}_{j}}$,
where $\rho_{\mathcal{Q}_{j}}$ is a density operator on
$\mathcal{H}_{\mathcal{Q}_{j}}$. If a state is not separable, then
it is called an entangled state.

\section{Positive operator valued measure on quantum phase }
In this section we will define a general POVM on quantum phase,
see Ref.  \cite{Hosh3}. This POVM is a set of linear operators
$\Delta(\varphi_{1,2},\ldots,\varphi_{1,N_{j}},\varphi_{2,3},\ldots,
\varphi_{N_{j}-1,N_{j}})$ furnishing the probabilities that the
measurement of a state $\rho_{Q_{j}}$ on the Hilbert space
$\mathcal{H}_{Q_{j}}$ is given by
\begin{eqnarray}&&p(\varphi_{1,2},\ldots,
\varphi_{1,N_{j}},\varphi_{2,3},\ldots,\varphi_{N_{j}-1,N_{j}})\\\nonumber&&=
\mathrm{Tr}(\rho\Delta
(\varphi_{1,2},\ldots,\varphi_{1,N_{j}},\varphi_{2,3},
\ldots,\varphi_{N_{j}-1,N_{j}})),
   \end{eqnarray}
where
$(\varphi_{1,2},\ldots,\varphi_{1,N_{j}},\varphi_{2,3},\ldots,\varphi_{N_{j}-1,N_{j}})$
are the outcomes of the measurement of the quantum  phase. This
POVM satisfies the following properties,
$\Delta(\varphi_{1,2},\ldots,\varphi_{1,N_{j}},\varphi_{2,3},\ldots,\varphi_{N_{j}-1,N_{j}})$
is self-adjoint, positive, and  normalized, that is
\begin{eqnarray}&&\overbrace{\int_{2\pi}\cdots
\int_{2\pi}}^{\frac{N_{j}(N_{j}-1)}{2}}d\varphi_{1,2}\cdots
d\varphi_{1,N_{j}}d\varphi_{2,3}\cdots
d\varphi_{N_{j}-1,N_{j}}\\\nonumber&&
    \Delta(\varphi_{1,2},\ldots,\varphi_{1,N_{j}},\varphi_{2,3},\ldots,\varphi_{N_{j}-1,N_{_{j}}})
    =\mathcal{I}_{N_{j}},
   \end{eqnarray}
   where
 the integral extends over any $2\pi$ intervals.
A general and symmetric POVM in a single $N_{j}$-dimensional
Hilbert space $\mathcal{H}_{\mathcal{Q}_{j}}$ is given by
\begin{eqnarray}
&&\Delta(\varphi_{1_j,2_j},\ldots,\varphi_{1_j,N_j},
\varphi_{2_j,3_j},\ldots,\varphi_{N_{j}-1,N_{j}})=
\sum^{N_{j}}_{l_{j},k_{j}=1}
e^{i\varphi_{k_{j},l_{j}}}\ket{k_{j}}\bra{l_{j}}\\\nonumber&&=
   \left(%
\begin{array}{ccccc}
  1 &e^{i\varphi_{1,2}}  & \cdots
  &  e^{i\varphi_{1,N_{j}-1}} &e^{i\varphi_{1,N_{j}}}\\
 e^{-i\varphi_{1,2}} &  1 & \cdots
 &  e^{i\varphi_{2,N_{j}-1}} &e^{i\varphi_{2,N_{j}}}\\
  \vdots&  \vdots&\ddots &\vdots& \vdots\\
    e^{-i\varphi_{1,N_{j}-1}} & e^{-i\varphi_{2,N_{j}-1}} &\cdots&1&e^{i\varphi_{N_{j}-1,N_{j}}}\\
  e^{-i\varphi_{1,N_{j}}} & e^{-i\varphi_{2,N_{j}}} &\cdots&e^{-i\varphi_{N_{j}-1,N_{j}}}&1\\
\end{array}%
\right),
\end{eqnarray}
where $\ket{k_{j}}$ and $\ket{l_{j}}$ are the basis vectors in
$\mathcal{H}_{\mathcal{Q}_j}$ and the quantum phases satisfy the
following relation $ \varphi_{k_{j},l_{j}}=
-\varphi_{l_{j},k_{j}}(1-\delta_{k_{j} l_{j}})$. The POVM is a
function of the $N_{j}(N_{j}-1)/2$ phases
$(\varphi_{1_j,2_j},\ldots,\varphi_{1_j,N_j},\varphi_{2_j,3_j},\ldots,\varphi_{N_{j}-1,N_{j}})$.
It is now possible to form a POVM of a multipartite system by
simply forming the tensor product
\begin{eqnarray}\label{POVM}\nonumber
\Delta_\mathcal{Q}(\varphi_{\mathcal{Q}_{1};k_{1},l_{1}},\ldots,
\varphi_{\mathcal{Q}_{m};k_{m},l_{m}})&=&
\Delta_{\mathcal{Q}_{1}}(\varphi_{\mathcal{Q}_{1};k_{1},l_{1}})
\otimes\cdots
\otimes\Delta_{\mathcal{Q}_{m}}(\varphi_{\mathcal{Q}_{m};k_{m},l_{m}}),
\\
\end{eqnarray}
where, e.g., $\varphi_{\mathcal{Q}_{1};k_{1},l_{1}}$ is the set of
POVMs phase associated with subsystems $\mathcal{Q}_{1}$, for all
$k_{1},l_{1}=1,2,\ldots,N_{1}$, where we need only to consider
when $l_{1}>k_{1}$.

\section{Concurrence for general pure bipartite states}
The concurrence of two-qubit states is defined as
  $\mathcal{C}(\Psi)=|\langle\Psi\ket{\widetilde{\Psi}}|$,
  where  the tilde represents the "spin-flip" operation
  $\ket{\widetilde{\Psi}}=\sigma_{y}\otimes
  \sigma_{y}\ket{\Psi^{*}}$, $\ket{\Psi^{*}}=\sum^{2}_{l,k=1}\alpha^{*}_{k,l}\ket{k,l}$ is the complex
  conjugate of $\ket{\Psi}=\sum^{2}_{l,k=1}\alpha_{k,l}\ket{k,l}$, and $\sigma_{y}=\left(%
\begin{array}{cc}
  0 & -i \\
  i & 0 \\
\end{array}%
\right)$ is a Pauli spin-flip operator
\cite{Wootters98,Wootters00}. Now, we will define concurrence for
a general pure bipartite state based on the orthogonal complement
of our POVM by constructing an antilinear operator for a general
pure bipartite state $\mathcal{Q}^{p}_{2}(N_{1},N_{2})$. The POVM
for quantum system $\mathcal{Q}^{p}_{2}(N_{1},N_{2})$ is given by
{\small\begin{eqnarray}
\Delta_\mathcal{Q}(\varphi_{\mathcal{Q}_{1};k_{1},l_{1}},
\varphi_{\mathcal{Q}_{2};k_{2},l_{2}})&=&
\Delta_{\mathcal{Q}_{1}}(\varphi_{\mathcal{Q}_{1};k_{1},l_{1}})
\otimes\Delta_{\mathcal{Q}_{2}}(\varphi_{\mathcal{Q}_{2};k_{2},l_{2}}).
\end{eqnarray}}
Next, we  define the orthogonal complement of our POVM by
$\widetilde{\Delta}_{\mathcal{Q}_{j}}(\varphi_{\mathcal{Q}_{j};k_{j},l_{j}})=\mathcal{I}_{N_{j}}-
\Delta_{\mathcal{Q}_{j}}(\varphi_{\mathcal{Q}_{j};k_{j},l_{j}})$,
where $\mathcal{I}_{N_{j}}$ is the $N_{j}$-by-$N_{j}$ identity
matrix, for each subsystem $j$. For example, for a  bipartite
state $\mathcal{Q}^{p}_{2}(2,3)$ we have
\begin{eqnarray}
 \widetilde{\Delta}_{\mathcal{Q}_{2}}(\varphi_{\mathcal{Q}_{2};k_{2},l_{2}})&=&
\widetilde{\Delta}_{\mathcal{Q}_{2}}(\varphi_{\mathcal{Q}_{2};1,2})+
\widetilde{\Delta}_{\mathcal{Q}_{2}}(\varphi_{\mathcal{Q}_{2};1,3})+
\widetilde{\Delta}_{\mathcal{Q}_{2}}(\varphi_{\mathcal{Q}_{2};2,3})
\\\nonumber&=&\left(%
\begin{array}{ccc}
  0 & e^{i\varphi_{\mathcal{Q}_{2};1,2}} & e^{i\varphi_{\mathcal{Q}_{2};1,3}} \\
  e^{-i\varphi_{\mathcal{Q}_{2};1,2}} & 0
  & e^{-i\varphi_{\mathcal{Q}_{2};2,3}} \\
  e^{-i\varphi_{\mathcal{Q}_{2};1,3}}
  & e^{-i\varphi_{\mathcal{Q}_{2};2,3}} & 0 \\
\end{array}%
\right)
\end{eqnarray}
where, e.g.,
$\widetilde{\Delta}_{\mathcal{Q}_{2}}(\varphi_{\mathcal{Q}_{2};1,3})
=\left(%
\begin{array}{ccc}
  0 & 0& e^{i\varphi_{\mathcal{Q}_{2};1,3}} \\
  0 & 0
  & 0 \\
  e^{-i\varphi_{\mathcal{Q}_{2};1,3}}
  & 0 & 0 \\
\end{array}%
\right)$. Moreover, we have
$\widetilde{\Delta}_{\mathcal{Q}_{1}}(\varphi_{\mathcal{Q}_{1};1,2})
=\left(%
\begin{array}{ccc}
  0 & e^{i\varphi_{\mathcal{Q}_{1};1,2}} \\
  e^{-i\varphi_{\mathcal{Q}_{1};1,2}} & 0 \\
\end{array}%
\right)$. Then for a quantum system $\mathcal{Q}^{p}_{2}(2,3)$ the
orthogonal complement of our POVM
$\Delta_\mathcal{Q}(\varphi_{\mathcal{Q}_{1};k_{1},l_{1}},
\varphi_{\mathcal{Q}_{2};k_{2},l_{2}})$ is given by
\begin{eqnarray}
\widetilde{\Delta}_\mathcal{Q}(\varphi_{\mathcal{Q}_{1};k_{1},l_{1}},
\varphi_{\mathcal{Q}_{2};k_{2},l_{2}})
&=&\widetilde{\Delta}_{\mathcal{Q}_{1}}
(\varphi_{\mathcal{Q}_{1};k_{1},l_{1}})\otimes
\widetilde{\Delta}_{\mathcal{Q}_{2}}
(\varphi_{\mathcal{Q}_{2};k_{2},l_{2}})\\\nonumber&=&
\widetilde{\Delta}_{\mathcal{Q}_{1}}
(\varphi_{\mathcal{Q}_{1};1,2})\otimes
\widetilde{\Delta}_{\mathcal{Q}_{2}}(\varphi_{\mathcal{Q}_{2};1,2})\\\nonumber&&+
\widetilde{\Delta}_{\mathcal{Q}_{1}}
(\varphi_{\mathcal{Q}_{1};1,2})\otimes\widetilde{\Delta}_{\mathcal{Q}_{2}}
(\varphi_{\mathcal{Q}_{2};1,3})\\\nonumber&&+
\widetilde{\Delta}_{\mathcal{Q}_{1}}
(\varphi_{\mathcal{Q}_{1};1,2})\otimes\widetilde{\Delta}_{\mathcal{Q}_{2}}
(\varphi_{\mathcal{Q}_{2};2,3})
\end{eqnarray}
Now, we will introduce the following notation
\begin{eqnarray}\label{EPRpm}
 \widetilde{\Delta}^{
EPR_{k_{1},l_{1};k_{2},l_{2}}}_{\mathcal{Q}_{1,2}(2,3)}&=&
\widetilde{\Delta}_{\mathcal{Q}_{1}}
(\varphi^{\frac{\pi}{2}}_{\mathcal{Q}_{1};1,2})
\otimes\widetilde{\Delta}_{\mathcal{Q}_{2}}
(\varphi^{\frac{\pi}{2}}_{\mathcal{Q}_{2};k_{2},l_{2}}),
\end{eqnarray}
where by choosing
$\varphi^{\frac{\pi}{2}}_{\mathcal{Q}_{j};k_{j},l_{j}}=\frac{\pi}{2}$
for all $k_{j}<l_{j}, ~j=1,2$, we get an operator which has the
structure of Pauli spin-flip operator $\sigma_{y}$ embedded in a
higher-dimensional Hilbert space and coincides with $\sigma_{y}$
for a single-qubit. Moreover, $EPR$ indicates that this operator
detects the generic bipartite entangled state. We then define the
 concurrence of quantum system $\mathcal{Q}^{p}_{2}(2,3)$  as
\begin{eqnarray}
    \mathcal{C}(\mathcal{Q}^{p}_{2}(2,3))&=&
    \left(\sum_{\forall k_{1},l_{1},k_{2},l_{2}}
    \left|\langle \Psi\ket{\widetilde{\Delta}^{
EPR_{k_{1},l_{1};k_{2},l_{2}}}_{\mathcal{Q}_{1,2}
(2,3)}\mathcal{C}_{2}\Psi}\right|^{^{2}}\right)^{1/2}
\\\nonumber&=&
 (
|\langle \Psi\ket{\widetilde{\Delta}^{
EPR_{1,2;1,2}}_{\mathcal{Q}_{1,2}
(2,3)}\mathcal{C}_{2}\Psi}|^{^{2}}+ |\langle
\Psi\ket{\widetilde{\Delta}^{EPR_{1,2;1,3}}_{\mathcal{Q}_{1,2}
(2,3)}\mathcal{C}_{2}\Psi}|^{^{2}}\\\nonumber&&+ |\langle
\Psi\ket{\widetilde{\Delta}^{EPR_{1,2;2,3}}_{\mathcal{Q}_{1,2}
(2,3)}\mathcal{C}_{2}\Psi}|^{^{2}})^{1/2}
\\\nonumber&=&
 ( 4\mathcal{N}^{EPR}_{2}[ |\alpha_{1,1}\alpha_{2,2}
-\alpha_{1,2}\alpha_{2,1}|^{2}+ |\alpha_{1,1}\alpha_{2,3}
-\alpha_{1,3}\alpha_{2,1}|^{2}
\\\nonumber&&+|\alpha_{1,2}\alpha_{2,3}
-\alpha_{1,3}\alpha_{2,2}|^{2}
 ])^{1/2} .
\end{eqnarray}
Now, the generalization of this result is straightforward. Hence,
for a pure quantum system $\mathcal{Q}^{p}_{2}(N_{1},N_{2})$ we
have
\begin{eqnarray}
 \widetilde{\Delta}^{
EPR_{k_{1},l_{1};k_{2},l_{2}}}_{\mathcal{Q}_{1,2}(N_{1},N_{2})}&=&\widetilde{\Delta}_{\mathcal{Q}_{1}}
(\varphi^{\frac{\pi}{2}}_{\mathcal{Q}_{1};k_{1},l_{1}})
\otimes\widetilde{\Delta}_{\mathcal{Q}_{2}}(\varphi^{\frac{\pi}{2}}_{\mathcal{Q}_{2};k_{2},l_{2}}),
\end{eqnarray}
and
 the concurrence is given by
\begin{eqnarray}
    \mathcal{C}(\mathcal{Q}^{p}_{2}(N_{1},N_{2}))&=&
    \left(\sum_{\forall k_{1},l_{1},k_{2},l_{2}}
    \left|\langle \Psi\ket{\widetilde{\Delta}^{
EPR_{k_{1},l_{1};k_{2},l_{2}}}_{\mathcal{Q}_{1,2}
(N_{1},N_{2})}\mathcal{C}_{2}\Psi}\right|^{^{2}}\right)^{1/2}\\\nonumber&=&
\left ( 4\mathcal{N}^{EPR}_{2} \sum^{N_{1}}_{l_{1}>k_{1}=1}
\sum^{N_{2}}_{l_{2}>k_{2}=1}
\left|\alpha_{k_{1},k_{2}}\alpha_{l_{1},l_{2}}-\alpha_{k_{1},l_{2}}\alpha_{l_{1},k_{2}}\right|^{2}
\right )^{1/2} .
\end{eqnarray}
The concurrence $ \mathcal{C}(\mathcal{Q}^{p}_{2}(N_{1},N_{2}))$
vanishes for product states, and coincides with our entanglement
tensor for general bipartite state \cite{Hosh4} and with the
generalized concurrence given in \cite{Gerjuoy,Albeverio}.
Moreover, the concurrence $
\mathcal{C}(\mathcal{Q}^{p}_{2}(N_{1},N_{2}))$ coincides with
I-concurrence, which is a generalization of concurrence introduced
by Rungta \emph{et al.} \cite{Rungta01} based on universal state
inversion. Furthermore, our antilinear operator
$\widetilde{\Delta}^{
EPR_{k_{1},l_{1};k_{2},l_{2}}}_{\mathcal{Q}_{1,2}
(N_{1},N_{2})}\mathcal{C}_{2}$ is invariant under LOCC operation
by construction.

\section{Concurrence classes for general pure multipartite  states}
In this section, we will construct concurrence classes for general
pure multipartite states
$\mathcal{Q}^{p}_{m}(N_{1},\ldots,N_{m})$. In order to simplify
our presentation,  we will use $\Lambda_{m}=k_{1},l_{1};$
$\ldots;k_{m},l_{m}$ as an abstract multi-index notation. The
unique structure of our POVM enables us to distinguish different
classes of multipartite states, which are inequivalent under LOCC
operations. In the $m$-partite case, the  off-diagonal elements of
the matrix corresponding to
\begin{eqnarray}\nonumber
\widetilde{\Delta}_\mathcal{Q}(\varphi_{\mathcal{Q}_{1};k_{1},l_{1}},\ldots,
\varphi_{\mathcal{Q}_{m};k_{m},l_{m}})&=&
\widetilde{\Delta}_{\mathcal{Q}_{1}}(\varphi_{\mathcal{Q}_{1};k_{1},l_{1}})
\otimes\cdots
\otimes\widetilde{\Delta}_{\mathcal{Q}_{m}}(\varphi_{\mathcal{Q}_{m};k_{m},l_{m}}),
\\
\end{eqnarray}
 have phases that are sum
or differences of phases originating from two and $m$ subsystems.
That is, in the later case the phases of
$\widetilde{\Delta}_\mathcal{Q}(\varphi_{\mathcal{Q}_{1};k_{1},l_{1}},\ldots,
\varphi_{\mathcal{Q}_{m};k_{m},l_{m}})$ take the form
$(\varphi_{\mathcal{Q}_{1};k_{1},l_{1}}\pm\varphi_{\mathcal{Q}_{2};k_{2},l_{2}}
\pm\ldots\pm\varphi_{\mathcal{Q}_{m};k_{m},l_{m}})$ and
identification of these joint phases makes our classification
possible. Thus, we  can define linear operators for  the $W^{m}$
class based on our POVM which are sum and difference of phases of
two subsystems, i.e.,
$(\varphi_{\mathcal{Q}_{r_{1}};k_{r_{1}},l_{r_{1}}}
\pm\varphi_{\mathcal{Q}_{r_{2}};k_{r_{2}},l_{r_{2}}})$. That is,
for the $W^{m}$ class we have
\begin{eqnarray}
 \widetilde{\Delta}^{
W^{m}_{\Lambda_{m}}}_{\mathcal{Q}_{r_{1},r_{2}}(N_{r_{1}},N_{r_{2}})}
&=&\mathcal{I}_{N_{1}} \otimes\cdots
\otimes\widetilde{\Delta}_{\mathcal{Q}_{r_{1}}}
(\varphi^{\frac{\pi}{2}}_{\mathcal{Q}_{r_{1}};k_{r_{1}},l_{r_{1}}})\\\nonumber&&
\otimes\cdots\otimes \widetilde{\Delta}_{\mathcal{Q}_{r_{2}}}
(\varphi^{\frac{\pi}{2}}_{\mathcal{Q}_{r_{2}};k_{r_{2}},l_{r_{2}}})\otimes\cdots\otimes
\mathcal{I}_{N_{m}}.
\end{eqnarray}
Let $C(m,k)=\left(%
\begin{array}{c}
  m \\
  k \\
\end{array}%
\right)$ denotes the binomial coefficient. Then there is $C(m,2)$
linear operators for the $W^{m}$ class and the set of these
operators gives the $W^{m}$ class concurrence.

 For the $GHZ^{m}$ class, we define the linear
operators based on our POVM which are sum and difference of phases
of $m$-subsystems, i.e.,
$(\varphi_{\mathcal{Q}_{r_{1}};k_{r_{1}},l_{r_{1}}}
\pm\varphi_{\mathcal{Q}_{r_{2}};k_{r_{2}},l_{r_{2}}}\pm
\ldots\pm\varphi_{\mathcal{Q}_{m};k_{m},l_{m}})$. That is, for the
$GHZ^{m}$ class we have
\begin{eqnarray}
 \widetilde{\Delta}^{
GHZ^{m}_{\Lambda_{m}}}_{\mathcal{Q}_{r_{1},r_{2}}(N_{r_{1}},N_{r_{2}})}
&=&\widetilde{\Delta}_{\mathcal{Q}_{1}}
(\varphi^{\pi}_{\mathcal{Q}_{1};k_{1},l_{1}})\otimes\cdots
\otimes\widetilde{\Delta}_{\mathcal{Q}_{r_{1}}}
(\varphi^{\frac{\pi}{2}}_{\mathcal{Q}_{r_{1}};k_{r_{1}},l_{r_{1}}})\\\nonumber&&
\otimes\cdots\otimes \widetilde{\Delta}_{\mathcal{Q}_{r_{2}}}
(\varphi^{\frac{\pi}{2}}_{\mathcal{Q}_{r_{2}};k_{r_{2}},l_{r_{2}}})\otimes\cdots\otimes
\widetilde{\Delta}_{\mathcal{Q}_{m}}
(\varphi^{\pi}_{\mathcal{Q}_{m};k_{m},l_{m}}).
\end{eqnarray}
where by choosing
$\varphi^{\pi}_{\mathcal{Q}_{j};k_{j},l_{j}}=\pi$ for all
$k_{j}<l_{j}, ~j=1,2,\ldots,m$, we get an operator which has the
structure of Pauli operator $\sigma_{x}$ embedded in a
higher-dimensional Hilbert space and coincides with $\sigma_{x}$
for a single-qubit. There are $C(m,2)$ linear operators for the
$GHZ^{m}$ class and the set of these operators gives the $GHZ^{m}$
class concurrence.

Moreover, we define the linear operators for the $GHZ^{m-1}$ class
of $m$-partite states based on our POVM which are sum and
difference of phases of $m-1$-subsystems, i.e.,
$(\varphi_{\mathcal{Q}_{r_{1}};k_{r_{1}},l_{r_{1}}}
\pm\varphi_{\mathcal{Q}_{r_{2}};k_{r_{2}},l_{r_{2}}}
\pm\ldots\varphi_{\mathcal{Q}_{m-1};k_{m-1},l_{m-1}}\pm\varphi_{\mathcal{Q}_{m-1};k_{m-1},l_{m-1}})$.
That is, for the $GHZ^{m-1}$ class we have

\begin{eqnarray}
\widetilde{\Delta}^{
GHZ^{m-1}_{\Lambda_{m}}}_{\mathcal{Q}_{r_{1}r_{2},r_{3}}(N_{r_{1}},N_{r_{2}})}
&=& \widetilde{\Delta}_{\mathcal{Q}_{r_{1}}}
(\varphi^{\frac{\pi}{2}}_{\mathcal{Q}_{r_{1}};k_{r_{1}},l_{r_{1}}})
\otimes\widetilde{\Delta}_{\mathcal{Q}_{r_{2}}}
(\varphi^{\frac{\pi}{2}}_{\mathcal{Q}_{r_{2}};k_{r_{2}},l_{r_{2}}})
\otimes\\\nonumber&&\widetilde{\Delta}_{\mathcal{Q}_{r_{3}}}
(\varphi^{\pi}_{\mathcal{Q}_{r_{3}};k_{r_{3}},l_{r_{3}}})
\otimes\cdots
\otimes\\\nonumber&&\widetilde{\Delta}_{\mathcal{Q}_{m-1}}
(\varphi^{\pi}_{\mathcal{Q}_{m-1};k_{r_{m-1}},l_{r_{m-1}}})\otimes\mathcal{I}_{N_{m}}
,
\end{eqnarray}
where $1\leq r_{1}<r_{2}<\cdots<r_{m-1}<m$. There is $C(m,m-1)$
such operators for the $GHZ^{m-1}$ class.

Now, we can construct concurrence classes for multipartite states.
For example, let $X^{m}$ denote two different classes of general
multipartite states, namely $W^{m}$ and $GHZ^{m}$ classes. Then,
for general pure quantum system
$\mathcal{Q}^{p}_{m}(N_{1},\ldots,N_{m})$ with
\begin{eqnarray}
   \mathcal{C}(\mathcal{Q}^{X^{m}}_{r_{1},r_{2}}(N_{r_{1}},N_{r_{2}}))&=&
    \sum_{\forall k_{1},l_{1},\ldots,k_{m},l_{m}}
    \left|\langle \Psi\ket{\widetilde{\Delta}^{
X^{m}_{\Lambda_{m}}}_{\mathcal{Q}_{r_{1},r_{2}}(N_{r_{1}},N_{r_{2}})}\mathcal{C}_{m}\Psi}
\right|^{^{2}},
\end{eqnarray}
the $X^{m}$ class concurrences are given by
\begin{eqnarray}
\mathcal{C}(\mathcal{Q}^{X^{m}}_{m}(N_{1},\ldots,N_{m}))&=&
    \left(\mathcal{N}^{X}_{m}\sum^{m}_{r_{2}>r_{1}=1}\mathcal{C}(\mathcal{Q}^{X^{m}}_{r_{1},r_{2}}
    (N_{r_{1}},N_{r_{2}}))\right)^{1/2},
\end{eqnarray}
where $\mathcal{N}^{X}_{m}$ is a normalization constant. Note that
for $m$-partite states the $W^{m}$ class concurrences are zero
only for completely separable states. That is as long as we have
bipartite entanglement in our state this measure does not vanish.
Thus, this class also  includes all biseparable states. We will
discuss this issue in detail in a forthcoming paper.  One can say
that the $W^{m}$ class concurrence measure the amount of bipartite
entanglement in a multipartite state.
 Now, let us address the monotonicity of these concurrence
classes of multipartite states. For $m$-qubit states, the $W^{m}$
class concurrences are entanglement monotones.  Let $A_{j}\in
SL(2,\mathbf{C})$, for $j=1,2,\ldots,m$, and
    $\mathcal{A}=A_{1}\otimes A_{2}\otimes\cdots\otimes A_{m}$, then
    $ \mathcal{A}\widetilde{\Delta}^{
W^{m}_{1,2;\ldots;1,2}}_{\mathcal{Q}_{r_{1},r_{2}}(2_{r_{1}},2_{r_{2}})}
\mathcal{A}^{T}=\widetilde{\Delta}^{
W^{m}_{1,2;\ldots;1,2}}_{\mathcal{Q}_{r_{1},r_{2}}(2_{r_{1}},2_{r_{2}})}$,
for all $1<r_{1}<r_{2}<m$. Thus, the $W^{m}$ class concurrences
for multi-qubit states are invariant under SLOCC, and hence are
entanglement monotones. Again, for general multipartite states we
cannot give any proof on invariance of $W^{m}$ class concurrence
under  SLOCC and this question needs further investigation.
Moreover, for multipartite states, the $GHZ^{m}$ class
concurrences are not entanglement monotone except under additional
conditions. Since
    $ \mathcal{A}\widetilde{\Delta}^{
GHZ^{m}_{1,2;\ldots;1,2}}_{\mathcal{Q}_{r_{1},r_{2}}(2_{r_{1}},2_{r_{2}})}
\mathcal{A}^{T}\neq\widetilde{\Delta}^{
GHZ^{3}_{1,2;\ldots;1,2}}_{\mathcal{Q}_{r_{1},r_{2}}(2_{r_{1}},2_{r_{2}})}$,
for all $1<r_{1}<r_{2}<m$. The reason is that $ A_{j}
\widetilde{\Delta}_{\mathcal{Q}_{j}}(\varphi^{\pi}_{\mathcal{Q}_{j};1,2})A^{T}_{j}
\neq\widetilde{\Delta}_{\mathcal{Q}_{j}}(\varphi^{\pi}_{\mathcal{Q}_{j};1,2})$.
Thus, the  $GHZ^{m}$ class concurrence for three-qubit states are
not invariant under SLOCC, and hence are not entanglement
monotones. However, by construction  the $GHZ^{m}$ class
concurrences are invariant under all permutations. Moreover, we
have $(\widetilde{\Delta}^{
GHZ^{3}_{1,2;\ldots;1,2}}_{\mathcal{Q}_{r_{1},r_{2}}(2_{r_{1}},2_{r_{2}})})^{2}=1$
and
$(\widetilde{\Delta}_{\mathcal{Q}_{j}}(\varphi^{\pi}_{\mathcal{Q}_{j};1,2}))^{2}=1$.
Furthermore, we need to be very careful when we are using the
$GHZ^{m}$ class concurrences. This class can be zero even for an
entangled multipartite state. Since we have more than two joint
phases in our POVM for $GHZ^{m}$ class concurrence. Thus, for the
$GHZ^{m}$ class concurrences we need to perform an optimization
over local unitary operations. For example, let
$\mathcal{U}=U_{1}\otimes U_{2}\otimes\cdots\otimes U_{m}$, where
$U_{j}\in U(N_{j},\mathbf{C})$. Then we maximize the $GHZ^{m}$
class concurrences for a given pure $m$-partite state over all
local unitary operations $\mathcal{U}$. If $\max_{\forall
\mathcal{U}}\mathcal{C}(\mathcal{Q}^{W^{m}}_{m}(N_{1},\ldots,N_{m}))\neq0$,
then we have a genuine $GHZ^{m}$ multipartite state by
construction. As an example of multi-qubit state let us consider a
state $\ket{W^{m}}=\frac{1}{\sqrt{m}}(\ket{1,1,\ldots,1,2}
+\ldots+\ket{2,1,\ldots,1,1})$. For this state the $W^{m}$ class
concurrence is
\begin{equation}
\mathcal{C}(\mathcal{Q}^{W^{m}}_{m}(2,\ldots,2))=
    (\frac{4C(m,2)}{m^{2}}\mathcal{N}^{W}_{m})^{1/2}
    =(\frac{ 2(m-1)}{m}\mathcal{N}^{W}_{m})^{1/2}.
\end{equation}
 This value coincides with the one given by D\"{u}r
\cite{Dur00}. Moreover, let us consider the following  state
$\ket{GHZ^{m}}=\frac{1}{\sqrt{2}}(\ket{1,\ldots,1}+\ket{2,\ldots,2})
$. For this state the $GHZ^{m}$ class concurrence is
\begin{equation}
\mathcal{C}(\mathcal{Q}^{GHZ^{m}}_{m}(2,\ldots,2))=
    (\frac{4C(m,2)}{4}\mathcal{N}^{GHZ}_{m})^{1/2}
    =(\frac{ m(m-1)}{2}\mathcal{N}^{GHZ}_{m})^{1/2}.
\end{equation}
We will discuss in detail these states for three-qubit and
four-qubit states below. Finally, for some partially separable
states the
$\mathcal{C}(\mathcal{Q}^{W^{m}}_{m}(N_{1},\ldots,N_{m}))$ class
and $\mathcal{C}(\mathcal{Q}^{GHZ^{m}}_{m}(N_{1},\ldots,N_{m}))$
class concurrences do not exactly quantify entanglement in
general. Example of such states can be e.g., constructed for
three-qubit states.


\section{Concurrence classes for general pure three-partite states}
In this section we will construct concurrences for  general pure
three-partite states based on orthogonal complement of our POVM.
 For three-partite states we have two
different joint phases in our POVM, those which are sum and
difference of phases of two subsystem, i.e.,
$(\varphi_{\mathcal{Q}_{1};k_{1},l_{1}}\pm\varphi_{\mathcal{Q}_{2};k_{2},l_{2}})$
and those which are sum and difference of phases of three
subsystem, i.e.,
$(\varphi_{\mathcal{Q}_{1};k_{1},l_{1}}\pm\varphi_{\mathcal{Q}_{2};k_{2},l_{2}}
\pm\varphi_{\mathcal{Q}_{3};k_{3},l_{3}})$. The first one
identifies $W^{3}$ class and the second one identifies  $GHZ^{3}$
class.  For $W^{3}$ class concurrence, we have three types of
entanglement, entanglement between subsystems one and two
$\mathcal{Q}_{1}\mathcal{Q}_{2}$, one and three
$\mathcal{Q}_{1}\mathcal{Q}_{3}$, and two and three
$\mathcal{Q}_{2}\mathcal{Q}_{3}$. So, we define a linear operator
by
\begin{eqnarray}\label{W3}\nonumber
 \widetilde{\Delta}^{
W^{3}_{\Lambda_{3}}}_{\mathcal{Q}_{1,2}(N_{1},N_{2})}
&=&\widetilde{\Delta}_{\mathcal{Q}_{1}}
(\varphi^{\frac{\pi}{2}}_{\mathcal{Q}_{1};k_{1},l_{1}})
\otimes\widetilde{\Delta}_{\mathcal{Q}_{2}}
(\varphi^{\frac{\pi}{2}}_{\mathcal{Q}_{2};k_{2},l_{2}})\otimes
\mathcal{I}_{N_{3}}.
\end{eqnarray}
$
 \widetilde{\Delta}^{
W^{3}_{\Lambda_{3}}}_{\mathcal{Q}_{1,3}(N_{1},N_{3})} $ and $
 \widetilde{\Delta}^{
W^{3}_{\Lambda_{3}}}_{\mathcal{Q}_{2,3}(N_{2},N_{3})} $ are
defined
 in the similar way.  Now, for pure quantum system
$\mathcal{Q}^{p}_{3}(N_{1},N_{2},N_{3})$ with
\begin{eqnarray}\label{conW3}
   \mathcal{C}(\mathcal{Q}^{W^{3}}_{r_{1},r_{2}}(N_{r_{1}},N_{r_{2}}))&=&
   \sum_{\forall k_{1},l_{1},k_{2},l_{2},;k_{3},l_{3}}
    \left|\langle \Psi\ket{\widetilde{\Delta}^{
W^{3}_{\Lambda_{3}}}_{\mathcal{Q}_{r_{1},r_{2}}(N_{r_{1}},N_{r_{2}})}
\mathcal{C}_{3}\Psi}\right|^{^{2}}
\end{eqnarray}
 $W^{3}$ class concurrence is given by
\begin{eqnarray}\nonumber
    &&\mathcal{C}(\mathcal{Q}^{W^{3}}_{3}(N_{1},N_{2},N_{3}))=
    \left(\mathcal{N}^{W}_{3}\sum^{3}_{1=r_{1}<r_{2}}
    \mathcal{C}(\mathcal{Q}^{W}_{r_{1},r_{2}}
    (N_{r_{1}},N_{r_{2}}))\right)^{1/2}=
\\\nonumber&&(4\mathcal{N}^{W}_{3}[\sum^{N_{1}}_{l_{1}>k_{1}=1}
\sum^{N_{2}}_{l_{2}>k_{2}=1} \left|\sum^{N_{3}}_{k_{3}=l_{3}=1}(
\alpha_{k_{1},l_{2},k_{3}}\alpha_{l_{1},k_{2},l_{3}}-\alpha_{k_{1},k_{2},k_{3}}\alpha_{l_{1},l_{2},l_{3}}
) \right|^{2}
\\\nonumber&&+\sum^{N_{1}}_{l_{1}>k_{1}=1}
\sum^{N_{3}}_{l_{3}>k_{3}=1}\left |
\sum^{N_{2}}_{k_{2}=l_{2}=1}(\alpha_{k_{1},k_{2},l_{3}}\alpha_{l_{1},l_{2},k_{3}}-\alpha_{k_{1},k_{2},k_{3}}\alpha_{l_{1},l_{2},l_{3}}
) \right|^{2}\\&&+ \sum^{N_{2}}_{l_{2}>k_{2}=1}
\sum^{N_{3}}_{l_{3}>k_{3}=1} \left |\sum^{N_{1}}_{k_{1}=l_{1}=1}(
\alpha_{k_{1},k_{2},l_{3}}\alpha_{l_{1},l_{2},k_{3}}-\alpha_{k_{1},k_{2},k_{3}}\alpha_{l_{1},l_{2},l_{3}}
) \right|^{2}])^{1/2},
\end{eqnarray}
where $\mathcal{N}^{W}_{3}$ is a normalization constant. By
construction the $W^{3}$ class  concurrence for three-partite
states vanishes for product states. Now, for a state
$\ket{\Psi_{W^{3}}}=\alpha_{1,1,2}\ket{1,1,2}+\alpha_{1,2,1}\ket{1,2,1}
+\alpha_{2,1,1}\ket{2,1,1}$, the $W^{3}$ class  concurrence gives
\begin{eqnarray}\nonumber
    &&\mathcal{C}(\mathcal{Q}^{W^{3}}_{3}(2,2,2))=(4\mathcal{N}^{W}_{3}
[ \left|\alpha_{1,2,1}\alpha_{2,1,1} \right|^{2}+\left |
\alpha_{1,1,2}\alpha_{2,1,1} \right|^{2}+\left
|\alpha_{1,1,2}\alpha_{1,2,1}
 \right|^{2}])^{1/2}.
\end{eqnarray}
When
    $\alpha_{1,1,2}=\alpha_{1,2,1}=\alpha_{2,1,1}=\frac{1}{\sqrt{3}}$, we get
    $\mathcal{C}(\mathcal{Q}^{W^{3}}_{3}(2,2,2))=
    (\frac{4}{3}\mathcal{N}^{W}_{3})^{1/2}$ and $\mathcal{C}(\mathcal{Q}^{GHZ^{3}}_{3}(2,2,2))=0$.
    Thus, for $\mathcal{N}^{W}_{3}=\frac{3}{4}$, we have
    $\mathcal{C}(\mathcal{Q}^{W^{3}}_{3}(2,2,2))=1$.

 The second class of three-partite
state that we would like to consider is the $GHZ^{3}$ class. For
this class, we have three types of entanglement, so there are
three linear operators. The first one is given by
\begin{eqnarray}\nonumber
\widetilde{\Delta}^{ GHZ^{3}_{\Lambda_{3}}}_{\mathcal{Q}_{1,2}
(N_{1},N_{2})}&=& \widetilde{\Delta}_{\mathcal{Q}_{1}}
(\varphi^{\frac{\pi}{2}}_{\mathcal{Q}_{1};k_{1},l_{1}})
\otimes\widetilde{\Delta}_{\mathcal{Q}_{2}}
(\varphi^{\frac{\pi}{2}}_{\mathcal{Q}_{2};k_{2},l_{2}})\otimes
\widetilde{\Delta}_{\mathcal{Q}_{3}}
(\varphi^{\pi}_{\mathcal{Q}_{3};k_{3},l_{3}}).
\end{eqnarray}
$
 \widetilde{\Delta}^{
GHZ^{3}_{\Lambda_{3}}}_{\mathcal{Q}_{1,3}(N_{1},N_{3})} $ and $
 \widetilde{\Delta}^{
GHZ^{3}_{\Lambda_{3}}}_{\mathcal{Q}_{2,3}(N_{2},N_{3})} $ are
defined in similar way. Now, for a pure quantum system
$\mathcal{Q}^{p}_{3}(N_{1},N_{2},N_{3})$ with
\begin{eqnarray}\label{conGHZ3}
   \mathcal{C}(\mathcal{Q}^{GHZ^{3}}_{r_{1},r_{2}}(N_{r_{1}},N_{r_{2}}))
   &=&\sum_{\forall k_{1}<l_{1},k_{2}<l_{2},;k_{3}<l_{3}}
    \left|\langle \Psi\ket{\widetilde{\Delta}^{
GHZ^{3}_{\Lambda_{3}}}_{\mathcal{Q}_{r_{1},r_{2}}(N_{r_{1}},N_{r_{2}})}
\mathcal{C}_{3}\Psi}\right|^{^{2}},
\end{eqnarray}
the $GHZ^{3}$ class concurrence for general pure three-partite
states is given by
\begin{eqnarray}\label{cone}
  && \mathcal{C}(\mathcal{Q}^{GHZ^{3}}_{3}(N_{1},N_{2},N_{3}))=
    \left(\mathcal{N}^{GHZ}_{3}
    \sum^{3}_{1=r_{1}<r_{2}}\mathcal{C}(\mathcal{Q}^{GHZ}_{r_{1},r_{2}}
    (N_{r_{1}},N_{r_{2}})\right)^{1/2}\\\nonumber&&=(4\mathcal{N}^{GHZ}_{3}[
    \sum^{N_{1}}_{l_{1}>k_{1}}\sum^{N_{1}}_{k_{1}=1}
\sum^{N_{2}}_{l_{2}>k_{2}}\sum^{N_{2}}_{k_{2}=1}
\sum^{N_{3}}_{l_{3}>k_{3}}\sum^{N_{3}}_{k_{3}=1}
\\\nonumber&& |
\alpha_{k_{1},l_{2},l_{3}}\alpha_{l_{1},k_{2},k_{3}}+
\alpha_{k_{1},l_{2},k_{3}}\alpha_{l_{1},k_{2},l_{3}}
-\alpha_{k_{1},k_{2},l_{3}}\alpha_{l_{1},l_{2},k_{3}}
-\alpha_{k_{1},k_{2},k_{3}}\alpha_{l_{1},l_{2},l_{3}}
|^{2}+\\\nonumber&&
| \alpha_{k_{1},l_{2},l_{3}}\alpha_{l_{1},k_{2},k_{3}}-
\alpha_{k_{1},l_{2},k_{3}}\alpha_{l_{1},k_{2},l_{3}}
+\alpha_{k_{1},k_{2},l_{3}}\alpha_{l_{1},l_{2},k_{3}}
-\alpha_{k_{1},k_{2},k_{3}}\alpha_{l_{1},l_{2},l_{3}}|^{2}+\\\nonumber&&|
-\alpha_{k_{1},l_{2},l_{3}}\alpha_{l_{1},k_{2},k_{3}}+
\alpha_{k_{1},l_{2},k_{3}}\alpha_{l_{1},k_{2},l_{3}}
+\alpha_{k_{1},k_{2},l_{3}}\alpha_{l_{1},l_{2},k_{3}}
-\alpha_{k_{1},k_{2},k_{3}}\alpha_{l_{1},l_{2},l_{3}}|^{2}])^{1/2},
\end{eqnarray}
where $\mathcal{N}^{GHZ}_{3}$ is a normalization constant. Now,
for the state
$\ket{\Psi_{GHZ^{3}}}=\alpha_{1,1,1}\ket{1,1,1}+\alpha_{2,2,2}\ket{2,2,2}
$, the $GHZ^{3}$ class  concurrence gives
$\mathcal{C}(\mathcal{Q}^{GHZ^{3}}_{3}(2,2,2))=
    (12\mathcal{N}^{GHZ}_{3}|\alpha_{1,1,1}\alpha_{2,2,2}
    |^{2})^{1/2}$ and for
    $\alpha_{1,1,1}=\alpha_{2,2,2}=\frac{1}{\sqrt{2}}$, we get
    $\mathcal{C}(\mathcal{Q}^{GHZ^{3}}_{3}(2,2,2))=
    (3\mathcal{N}^{GHZ}_{3})^{1/2}$. Thus, for $\mathcal{N}^{GHZ}_{3}=\frac{1}{3}$ we have
    $\mathcal{C}(\mathcal{Q}^{GHZ^{3}}_{3}(2,2,2))=1$. However, for this state
    $\mathcal{C}(\mathcal{Q}^{W^{3}}_{3}(2,2,2))=0$.
    Note that for  some states we need  to perform  optimization over
    local unitary operations as was discussed in the previous
    section.

\section{Concurrence classes for general pure four-partite states
}\label{Sec4} For general four-partite states we have three
different joint phases in our POVM. Those which are sum and
difference of phases of two subsystems, i.e.,
$(\varphi_{\mathcal{Q}_{1};k_{1},l_{1}}\pm\varphi_{\mathcal{Q}_{2};k_{2},l_{2}})$
, those which are sum and difference of phases of three
subsystems, i.e.,
$(\varphi_{\mathcal{Q}_{1};k_{1},l_{1}}\pm\varphi_{\mathcal{Q}_{2};k_{2},l_{2}}
\pm\varphi_{\mathcal{Q}_{3};k_{3},l_{3}})$, and those which are
sum and difference of phases of four subsystems, i.e.,
$(\varphi_{\mathcal{Q}_{1};k_{1},l_{1}}\pm\varphi_{\mathcal{Q}_{2};k_{2},l_{2}}
\pm\varphi_{\mathcal{Q}_{3};k_{3},l_{3}}\pm\varphi_{\mathcal{Q}_{4};k_{4},l_{4}})$.
The first one identifies $W^{4}$ class concurrence, the second one
identifies $GHZ^{3}$ class concurrence, and the third one
identifies $GHZ^{4}$ class concurrence.
 For the $W^{4}$ class, we have six types
of entanglement, so there are six  operators corresponding to
entanglement between $\mathcal{Q}_{1}\mathcal{Q}_{2}$,
$\mathcal{Q}_{1}\mathcal{Q}_{3}$,
$\mathcal{Q}_{1}\mathcal{Q}_{4}$,
$\mathcal{Q}_{2}\mathcal{Q}_{3}$,
$\mathcal{Q}_{2}\mathcal{Q}_{4}$, and
$\mathcal{Q}_{3}\mathcal{Q}_{4}$ subsystems. The linear operator
corresponding to $\mathcal{Q}_{1}\mathcal{Q}_{2}$ is given by
\begin{eqnarray}\nonumber
 \widetilde{\Delta}^{
W^{4}_{\Lambda_{4}}}_{\mathcal{Q}_{1,2}(N_{1},N_{2})}&=&\widetilde{\Delta}_{\mathcal{Q}_{1}}
(\varphi^{\frac{\pi}{2}}_{\mathcal{Q}_{1};k_{1},l_{1}})
\otimes\widetilde{\Delta}_{\mathcal{Q}_{2}}
(\varphi^{\frac{\pi}{2}}_{\mathcal{Q}_{2};k_{2},l_{2}})
\otimes\mathcal{I}_{N_{3}}\otimes\mathcal{I}_{N_{4}}.
\end{eqnarray}
$
 \widetilde{\Delta}^{
W^{4}_{\Lambda_{4}}}_{\mathcal{Q}_{1,3}(N_{1},N_{3})} $, $
 \widetilde{\Delta}^{
W^{4}_{\Lambda_{4}}}_{\mathcal{Q}_{1,4}(N_{1},N_{4})}$, $
 \widetilde{\Delta}^{
W^{4}_{\Lambda_{4}}}_{\mathcal{Q}_{2,3}(N_{2},N_{3})} $, $
 \widetilde{\Delta}^{
W^{4}_{\Lambda_{4}}}_{\mathcal{Q}_{2,4}(N_{2},N_{4})} $, and $
 \widetilde{\Delta}^{
W^{4}_{\Lambda_{4}}}_{\mathcal{Q}_{3,4}(N_{2},N_{4})}$
 are defined in similar way.
Now, for a pure quantum system
$\mathcal{Q}^{p}_{4}(N_{1},\ldots,N_{4})$ with
\begin{eqnarray}
\mathcal{C}(\mathcal{Q}^{W^{4}}_{r_{1},r_{2}}(N_{r_{1}},N_{r_{2}}))&=&
\sum_{\forall k_{1},l_{1},\ldots,k_{4},l_{4}}
    \left|\langle \Psi\ket{\widetilde{\Delta}^{
W^{4}_{\Lambda_{4}}}_{\mathcal{Q}_{r_{1},r_{2}}(N_{r_{1}},N_{r_{2}})}
\mathcal{C}_{4}\Psi}\right|^{^{2}},
\end{eqnarray}
 the $W^{4}$ class  concurrence is given by
\begin{eqnarray}
    &&\mathcal{C}(\mathcal{Q}^{W^{4}}_{4}(N_{1},\ldots,N_{4}))=
    \left(\mathcal{N}^{W}_{4}\sum^{4}_{1=r_{1}<r_{2}}
    \mathcal{C}(\mathcal{Q}^{W^{4}}_{r_{1},r_{2}}
    (N_{r_{1}},N_{r_{2}}))\right)^{1/2}=(4\mathcal{N}^{W}_{4}\\\nonumber&&
    \sum^{N_{1}}_{l_{1}>k_{1}=1}
\sum^{N_{2}}_{l_{2}>k_{2}=1}
\left|\sum^{N_{3}}_{k_{3}=l_{3}=1}\sum^{N_{4}}_{k_{4}=l_{4}=1}
(\alpha_{k_{1},l_{2},k_{3},k_{4}}\alpha_{l_{1},k_{2},l_{3},l_{4}}-
\alpha_{k_{1},k_{2},k_{3},k_{4}}\alpha_{l_{1},l_{2},l_{3},l_{4}} )
\right|^{2}+\\\nonumber&& \sum^{N_{1}}_{l_{1}>k_{1}=1}
\sum^{N_{3}}_{l_{3}>k_{3}=1}
\left|\sum^{N_{2}}_{k_{2}=l_{2}=1}\sum^{N_{4}}_{k_{4}=l_{4}=1}
(\alpha_{k_{1},k_{2},l_{3},k_{4}}\alpha_{l_{1},l_{2},k_{3},l_{4}}
-\alpha_{k_{1},k_{2},k_{3},k_{4}}\alpha_{l_{1},l_{2},l_{3},l_{4}}
) \right|^{2}+\\\nonumber&&\sum^{N_{1}}_{l_{1}>k_{1}=1}
\sum^{N_{4}}_{l_{4}>k_{4}=1}
\left|\sum^{N_{2}}_{k_{2}=l_{2}=1}\sum^{N_{3}}_{k_{3}=l_{3}=1}
(\alpha_{k_{1},k_{2},k_{3},l_{4}}\alpha_{l_{1},l_{2},l_{3},k_{4}}-
\alpha_{k_{1},k_{2},k_{3},k_{4}}\alpha_{l_{1},l_{2},l_{3},l_{4}})
\right|^{2}+\\\nonumber&& \sum^{N_{2}}_{l_{2}>k_{2}=1}
\sum^{N_{3}}_{l_{3}>k_{3}=1}
\left|\sum^{N_{1}}_{k_{1}=l_{1}=1}\sum^{N_{4}}_{k_{4}=l_{4}=1}
(\alpha_{k_{1},k_{2},l_{3},k_{4}}\alpha_{l_{1},l_{2},k_{3},l_{4}}
-\alpha_{k_{1},k_{2},k_{3},k_{4}}\alpha_{l_{1},l_{2},l_{3},l_{4}})
\right|^{2}+\\\nonumber&& \sum^{N_{2}}_{l_{2}>k_{2}=1}
\sum^{N_{4}}_{l_{4}>k_{4}=1} \left
|\sum^{N_{1}}_{k_{1}=l_{1}=1}\sum^{N_{3}}_{k_{3}=l_{3}=1}
(\alpha_{k_{1},k_{2},k_{3},l_{4}}\alpha_{l_{1},l_{2},l_{3},k_{4}}-
\alpha_{k_{1},k_{2},k_{3},k_{4}}\alpha_{l_{1},l_{2},l_{3},l_{4}})
\right|^{2}+\\\nonumber&& \sum^{N_{3}}_{l_{3}>k_{3}=1}
\sum^{N_{4}}_{l_{4}>k_{4}=1}
\left|\sum^{N_{1}}_{k_{1}=l_{1}=1}\sum^{N_{2}}_{k_{2}=l_{2}=1}
(\alpha_{k_{1},k_{2},k_{3},l_{4}}\alpha_{l_{1},l_{2},l_{3},k_{4}}
-\alpha_{k_{1},k_{2},k_{3},k_{4}}\alpha_{l_{1},l_{2},l_{3},l_{4}}
)\right |^{2})^{1/2},
\end{eqnarray}
where $\mathcal{N}^{W}_{4}$ is a normalization constant. Now, for
a state
$\ket{\Psi_{W^{4}}}=\alpha_{1,1,1,2}\ket{1,1,1,2}+\alpha_{1,1,2,1}\ket{1,1,2,1}
+\alpha_{1,2,1,1}\ket{1,2,1,1}+\alpha_{2,1,1,1}\ket{2,1,1,1}$, the
$W^{4}$ class  concurrence
 gives
\begin{eqnarray}\nonumber
\mathcal{C}(\mathcal{Q}^{W^{4}}_{4}(2,\ldots,2))&=&(4\mathcal{N}^{W}_{3}
[ \left|\alpha_{1,2,1,1}\alpha_{2,1,1,1} \right|^{2}+\left |
\alpha_{1,1,2,1}\alpha_{2,1,1,1} \right|^{2}\\\nonumber&&+\left
|\alpha_{1,1,1,2}\alpha_{2,1,1,1}
 \right|^{2}+\left
|\alpha_{1,1,2,1}\alpha_{1,2,1,1}
 \right|^{2}\\\nonumber&&+\left
|\alpha_{1,1,1,2}\alpha_{1,2,1,1}
 \right|^{2}+\left
|\alpha_{1,1,1,2}\alpha_{1,1,2,1}
 \right|^{2}]
 )^{1/2}
\end{eqnarray} and for
    $\alpha_{1,1,1,2}=\alpha_{1,1,2,1}=\alpha_{1,2,1,1}=\alpha_{1,2,1,1}
    =\frac{1}{\sqrt{4}}$, we get
    $\mathcal{C}(\mathcal{Q}^{W^{4}}_{4}(2,\ldots,2))=
    (\frac{4C(4,2)}{4^{2}}\mathcal{N}^{W}_{4})^{1/2}=
    (\frac{3}{2}\mathcal{N}^{W}_{4})^{1/2}$,
    $\mathcal{C}(\mathcal{Q}^{GHZ^{3}}_{4}(2,\ldots,2))=0$.

The second class of four-partite state that we want to consider is
the $GHZ^{3}$ class. For this class, we have four types of
entanglement. These linear operators are given by
\begin{eqnarray}\nonumber
 \widetilde{\Delta}^{
GHZ^{3}_{\Lambda_{4}}}_{\mathcal{Q}_{12,3}(N_{1},N_{2})}&=&
\widetilde{\Delta}_{\mathcal{Q}_{1}}
(\varphi^{\frac{\pi}{2}}_{\mathcal{Q}_{1};k_{1},l_{1}})
\otimes\widetilde{\Delta}_{\mathcal{Q}_{2}}
(\varphi^{\frac{\pi}{2}}_{\mathcal{Q}_{2};k_{2},l_{2}})
\otimes\widetilde{\Delta}_{\mathcal{Q}_{3}}(\varphi^{\pi}_{\mathcal{Q}_{3};k_{3},l_{3}})
\otimes\mathcal{I}_{N_{4}},
\end{eqnarray}
\begin{eqnarray}\nonumber
 \widetilde{\Delta}^{
GHZ^{3}_{\Lambda_{4}}}_{\mathcal{Q}_{12,4}(N_{1},N_{3})}&=&
\widetilde{\Delta}_{\mathcal{Q}_{1}}
(\varphi^{\frac{\pi}{2}}_{\mathcal{Q}_{1};k_{1},l_{1}}) \otimes
\widetilde{\Delta}_{\mathcal{Q}_{2}}
(\varphi^{\frac{\pi}{2}}_{\mathcal{Q}_{2};k_{2},l_{2}})
\otimes\nonumber\mathcal{I}_{N_{3}}
\otimes\widetilde{\Delta}_{\mathcal{Q}_{4}}(\varphi^{\pi}_{\mathcal{Q}_{4};k_{4},l_{4}}),
\end{eqnarray}
\begin{eqnarray}\nonumber
 \widetilde{\Delta}^{
GHZ^{3}_{\Lambda_{4}}}_{\mathcal{Q}_{13,4}(N_{1},N_{4})}&=&
\widetilde{\Delta}_{\mathcal{Q}_{1}}
(\varphi^{\frac{\pi}{2}}_{\mathcal{Q}_{1};k_{1},l_{1}}) \otimes
\mathcal{I}_{N_{2}}
\otimes\nonumber\widetilde{\Delta}_{\mathcal{Q}_{3}}
(\varphi^{\frac{\pi}{2}}_{\mathcal{Q}_{3};k_{3},l_{3}})\otimes\widetilde{\Delta}_{\mathcal{Q}_{4}}
(\varphi^{\pi}_{\mathcal{Q}_{4};k_{4},l_{4}}),
\end{eqnarray}
 \begin{eqnarray}
 \widetilde{\Delta}^{
GHZ^{3}_{\Lambda_{4}}}_{\mathcal{Q}_{23,4}(N_{2},N_{3})}
&=&\mathcal{I}_{N_{1}} \otimes\widetilde{\Delta}_{\mathcal{Q}_{2}}
(\varphi^{\frac{\pi}{2}}_{\mathcal{Q}_{2};k_{2},l_{2}})
\otimes\nonumber \widetilde{\Delta}_{\mathcal{Q}_{3}}
(\varphi^{\frac{\pi}{2}}_{\mathcal{Q}_{3};k_{3},l_{3}})\otimes
\widetilde{\Delta}_{\mathcal{Q}_{4}}(\varphi^{\pi}_{\mathcal{Q}_{4};k_{4},l_{4}}),
\end{eqnarray}
where, e.g., $ \widetilde{\Delta}^{
GHZ^{3}_{\Lambda_{4}}}_{\mathcal{Q}_{12,4}(N_{1},N_{2})}$
identifies elements of our POVM  which are sum and difference of
phases of three subsystems, i.e.,
$([\varphi_{\mathcal{Q}_{1};k_{1},l_{1}}\pm\varphi_{\mathcal{Q}_{2};k_{2},l_{2}}]
\pm\varphi_{\mathcal{Q}_{4};k_{4},l_{4}})$ and extract information
about entanglement of subsystems
$\mathcal{Q}_{1}\mathcal{Q}_{2}-\mathcal{Q}_{4}$. For pure quantum
system $\mathcal{Q}^{p}_{4}(N_{1},\ldots,N_{4})$ with
\begin{eqnarray}
   \mathcal{C}(\mathcal{Q}^{GHZ^{3}}_{r_{1}r_{2},r_{3}}(N_{r_{1}},N_{r_{2}}))&=&
    \sum_{\forall k_{1},l_{1},\ldots,k_{4},l_{4}}
    \left|\langle \Psi\ket{\widetilde{\Delta}^{
GHZ^{3}_{\Lambda_{4}}}_{\mathcal{Q}_{r_{1}r_{2},r_{3}}
(N_{r_{1}},N_{r_{2}})}\mathcal{C}_{4}\Psi}\right|^{^{2}},
\end{eqnarray}
the $GHZ^{3}$ class concurrence for general four-partite state is
given by
\begin{eqnarray}\nonumber
   && \mathcal{C}(\mathcal{Q}^{GHZ^{3}}_{4}(N_{1},\ldots,N_{4}))=
    \left(\mathcal{N}^{GHZ}_{3}\sum^{4}_{1=r_{1}<r_{2}<r_{3}}
    \mathcal{C}(\mathcal{Q}^{GHZ^{3}_{\Lambda_{4}}}_{r_{1}r_{2},r_{3}}
    (N_{r_{1}},N_{r_{2}}))\right)^{1/2}\\\nonumber&&
    =(\mathcal{N}^{GHZ}_{3}\sum^{N_{1}}_{l_{1}>k_{1}=1}
\sum^{N_{2}}_{l_{2}>k_{2}=1} \sum^{N_{3}}_{l_{3}>k_{3}=1}
|\sum^{N_{4}}_{k_{4}=l_{4}=1}
(-\alpha_{k_{1},k_{2},k_{3},k_{4}}\alpha_{l_{1},l_{2},l_{3},l_{4}}
\\\nonumber&&
-\alpha_{k_{1},k_{2},l_{3},k_{4}}\alpha_{l_{1},l_{2},k_{3},l_{4}}
+\alpha_{k_{1},l_{2},k_{3},k_{4}}\alpha_{l_{1},k_{2},l_{3},l_{4}}
+\alpha_{k_{1},l_{2},l_{3},k_{4}}\alpha_{l_{1},k_{2},k_{3},l_{4}}
)|^{2}+\\\nonumber&&\sum^{N_{1}}_{l_{1}>k_{1}=1}
\sum^{N_{2}}_{l_{2}>k_{2}=1} \sum^{N_{4}}_{l_{4}>k_{4}=1}
|\sum^{N_{3}}_{k_{3}=l_{3}=1}(
-\alpha_{k_{1},k_{2},k_{3},k_{4}}\alpha_{l_{1},l_{2},l_{3},l_{4}}
-\alpha_{k_{1},k_{2},k_{3},l_{4}}\alpha_{l_{1},l_{2},l_{3},k_{4}}
\\\nonumber&&+
\alpha_{k_{1},l_{2},k_{3},k_{4}}\alpha_{l_{1},k_{2},l_{3},l_{4}}+
\alpha_{k_{1},l_{2},k_{3},l_{4}}\alpha_{l_{1},k_{2},l_{3},k_{4}}
)|^{2} +\\\nonumber&&\sum^{N_{1}}_{l_{1}>k_{1}=1}
\sum^{N_{3}}_{l_{3}>k_{3}=1} \sum^{N_{4}}_{l_{4}>k_{4}=1}
|\sum^{N_{2}}_{k_{2}=l_{2}=1} (
-\alpha_{k_{1},k_{2},k_{3},k_{4}}\alpha_{l_{1},l_{2},l_{3},l_{4}}-
\alpha_{k_{1},k_{2},k_{3},l_{4}}\alpha_{l_{1},l_{2},l_{3},k_{4}}
\\\nonumber&&+
\alpha_{k_{1},k_{2},l_{3},k_{4}}\alpha_{l_{1},l_{2},k_{3},l_{4}}+
\alpha_{k_{1},k_{2},l_{3},l_{4}}\alpha_{l_{1},l_{2},k_{3},k_{4}}
)|^{2}+\\\nonumber&&\sum^{N_{2}}_{l_{2}>k_{2}=1}
 \sum^{N_{3}}_{l_{3}>k_{3}=1}
\sum^{N_{4}}_{l_{4}>k_{4}=1}
 |\sum^{N_{1}}_{k_{1}=l_{1}=1}(
-\alpha_{k_{1},k_{2},k_{3},k_{4}}\alpha_{l_{1},l_{2},l_{3},l_{4}}+
\alpha_{k_{1},k_{2},k_{3},l_{4}}\alpha_{l_{1},l_{2},l_{3},k_{4}}
\\\nonumber
&& -
\alpha_{k_{1},k_{2},l_{3},k_{4}}\alpha_{l_{1},l_{2},k_{3},l_{4}}+
\alpha_{k_{1},k_{2},l_{3},l_{4}}\alpha_{l_{1},l_{2},k_{3},k_{4}}
|^{2})^{1/2},
\end{eqnarray}
where $\mathcal{N}^{GHZ}_{3}$ is a normalization constant.
 Next we are going to consider the $GHZ^{4}$class
 concurrence for general four-partite states. For the $GHZ^{4}$ class, we have again six types
of entanglement, so there are six linear operators corresponding
to entanglement between these subsystems. The linear operator
corresponding to
$(\mathcal{Q}_{1}\mathcal{Q}_{2})\mathcal{Q}_{3}\mathcal{Q}_{4}$
is given by
\begin{eqnarray}
 \widetilde{\Delta}^{
GHZ^{4}_{\Lambda_{4}}}_{\mathcal{Q}_{1,2}(N_{1},N_{2})}&=&
\widetilde{\Delta}_{\mathcal{Q}_{1}}
(\varphi^{\frac{\pi}{2}}_{\mathcal{Q}_{1};k_{1},l_{1}})
\otimes\widetilde{\Delta}_{\mathcal{Q}_{2}}
(\varphi^{\frac{\pi}{2}}_{\mathcal{Q}_{2};k_{2},l_{2}})
\\&\otimes&\nonumber\widetilde{\Delta}_{\mathcal{Q}_{3}}(\varphi^{\pi}_{\mathcal{Q}_{3};k_{3},l_{3}})
\otimes\widetilde{\Delta}_{\mathcal{Q}_{4}}(\varphi^{\pi}_{\mathcal{Q}_{4};k_{4},l_{4}}),
\end{eqnarray}
$
 \widetilde{\Delta}^{
GHZ^{4}_{\Lambda_{4}}}_{\mathcal{Q}_{1,3}(N_{1},N_{3})},~
 \widetilde{\Delta}^{
GHZ^{4}_{\Lambda_{4}}}_{\mathcal{Q}_{1,4}(N_{1},N_{4})},~
 \widetilde{\Delta}^{
GHZ^{4}_{\Lambda_{4}}}_{\mathcal{Q}_{2,3}(N_{2},N_{3})}, $
$\widetilde{\Delta}^{
GHZ^{4}_{\Lambda_{4}}}_{\mathcal{Q}_{2,4}(N_{2},N_{4})}$, and $
 \widetilde{\Delta}^{
GHZ^{4}_{\Lambda_{4}}}_{\mathcal{Q}_{3,4}(N_{3},N_{4})} $ are
defined in a similar way. Now, for a pure quantum system
$\mathcal{Q}^{p}_{4}(N_{1},\ldots,N_{4})$, let
\begin{eqnarray}
   \mathcal{C}(\mathcal{Q}^{GHZ^{4}}_{r_{1},r_{2}}(N_{r_{1}},N_{r_{2}}))&=&
    \sum_{\forall k_{1}<l_{1},\ldots,k_{4}<l_{4}}
    \left|\langle \Psi\ket{\widetilde{\Delta}^{
GHZ^{4}_{\Lambda_{4}}}_{\mathcal{Q}_{r_{1},r_{2}}
(N_{r_{1}},N_{r_{2}})}\mathcal{C}_{4}\Psi}\right|^{^{2}}.
\end{eqnarray}
Then, the $GHZ^{4}$ class concurrence is given by
\begin{eqnarray}
  && \mathcal{C}(\mathcal{Q}^{GHZ^{4}}_{4}(N_{1},\ldots,N_{4}))=
    \left(\mathcal{N}^{GHZ}_{4}\sum^{4}_{1=r_{1}<r_{2}}\mathcal{C}(\mathcal{Q}^{GHZ^{4}}_{r_{1},r_{2}}
    (N_{r_{1}},N_{r_{2}}))\right)^{1/2}
    \\\nonumber&&=(4\mathcal{N}^{GHZ}_{4}
    \sum^{N_{1}}_{l_{1}>k_{1}=1}
\sum^{N_{2}}_{l_{2}>k_{2}=1} \sum^{N_{3}}_{l_{3}>k_{3}=1}
\sum^{N_{4}}_{l_{4}>k_{4}=1} [
|-\alpha_{k_{1},k_{2},k_{3},k_{4}}\alpha_{l_{1},l_{2},l_{3},l_{4}}\\\nonumber
&&-
\alpha_{k_{1},k_{2},k_{3},l_{4}}\alpha_{l_{1},l_{2},l_{3},k_{4}}-
\alpha_{k_{1},k_{2},l_{3},k_{4}}\alpha_{l_{1},l_{2},k_{3},l_{4}}-
\alpha_{k_{1},k_{2},l_{3},l_{4}}\alpha_{l_{1},l_{2},k_{3},k_{4}}\\\nonumber
&&
+\alpha_{k_{1},l_{2},k_{3},k_{4}}\alpha_{l_{1},k_{2},l_{3},l_{4}}+
\alpha_{k_{1},l_{2},k_{3},l_{4}}\alpha_{l_{1},k_{2},l_{3},k_{4}}+
\alpha_{k_{1},l_{2},l_{3},k_{4}}\alpha_{l_{1},k_{2},k_{3},l_{4}}\\\nonumber
&&+
\alpha_{k_{1},l_{2},l_{3},l_{4}}\alpha_{l_{1},k_{2},k_{3},k_{4}}
|^{2} 
+|-\alpha_{k_{1},k_{2},k_{3},k_{4}}\alpha_{l_{1},l_{2},l_{3},l_{4}}
-\alpha_{k_{1},k_{2},k_{3},l_{4}}\alpha_{l_{1},l_{2},l_{3},k_{4}}\\\nonumber
&&+\alpha_{k_{1},k_{2},l_{3},k_{4}}\alpha_{l_{1},l_{2},k_{3},l_{4}}
+\alpha_{k_{1},k_{2},l_{3},l_{4}}\alpha_{l_{1},l_{2},k_{3},k_{4}}
-\alpha_{k_{1},l_{2},k_{3},k_{4}}\alpha_{l_{1},k_{2},l_{3},l_{4}}\\\nonumber
&&-
\alpha_{k_{1},l_{2},k_{3},l_{4}}\alpha_{l_{1},k_{2},l_{3},k_{4}}+
\alpha_{k_{1},l_{2},l_{3},k_{4}}\alpha_{l_{1},k_{2},k_{3},l_{4}}+
\alpha_{k_{1},l_{2},l_{3},l_{4}}\alpha_{l_{1},k_{2},k_{3},k_{4}}
|^{2}
\\\nonumber&&+
|-\alpha_{k_{1},k_{2},k_{3},k_{4}}\alpha_{l_{1},l_{2},l_{3},l_{4}}+
\alpha_{k_{1},k_{2},k_{3},l_{4}}\alpha_{l_{1},l_{2},l_{3},k_{4}}-
\alpha_{k_{1},k_{2},l_{3},k_{4}}\alpha_{l_{1},l_{2},k_{3},l_{4}}\\\nonumber
&&+
\alpha_{k_{1},k_{2},l_{3},l_{4}}\alpha_{l_{1},l_{2},k_{3},k_{4}}
-\alpha_{k_{1},l_{2},k_{3},k_{4}}\alpha_{l_{1},k_{2},l_{3},l_{4}}+
\alpha_{k_{1},l_{2},k_{3},l_{4}}\alpha_{l_{1},k_{2},l_{3},k_{4}}\\\nonumber
&&+
\alpha_{k_{1},l_{2},l_{3},k_{4}}\alpha_{l_{1},k_{2},k_{3},l_{4}}-
\alpha_{k_{1},l_{2},l_{3},l_{4}}\alpha_{l_{1},k_{2},k_{3},k_{4}}
|^{2}+
|-\alpha_{k_{1},k_{2},k_{3},k_{4}}\alpha_{l_{1},l_{2},l_{3},l_{4}}\\\nonumber
&&-
\alpha_{k_{1},k_{2},k_{3},l_{4}}\alpha_{l_{1},l_{2},l_{3},k_{4}}+
\alpha_{k_{1},k_{2},l_{3},k_{4}}\alpha_{l_{1},l_{2},k_{3},l_{4}}+
\alpha_{k_{1},k_{2},l_{3},l_{4}}\alpha_{l_{1},l_{2},k_{3},k_{4}}\\\nonumber
&&
+\alpha_{k_{1},l_{2},k_{3},k_{4}}\alpha_{l_{1},k_{2},l_{3},l_{4}}+
\alpha_{k_{1},l_{2},k_{3},l_{4}}\alpha_{l_{1},k_{2},l_{3},k_{4}}-
\alpha_{k_{1},l_{2},l_{3},k_{4}}\alpha_{l_{1},k_{2},k_{3},l_{4}}\\\nonumber
&&-
\alpha_{k_{1},l_{2},l_{3},l_{4}}\alpha_{l_{1},k_{2},k_{3},k_{4}}
|^{2} 
+|-\alpha_{k_{1},k_{2},k_{3},k_{4}}\alpha_{l_{1},l_{2},l_{3},l_{4}}+
\alpha_{k_{1},k_{2},k_{3},l_{4}}\alpha_{l_{1},l_{2},l_{3},k_{4}}\\\nonumber
&&-
\alpha_{k_{1},k_{2},l_{3},k_{4}}\alpha_{l_{1},l_{2},k_{3},l_{4}}+
\alpha_{k_{1},k_{2},l_{3},l_{4}}\alpha_{l_{1},l_{2},k_{3},k_{4}}
+\alpha_{k_{1},l_{2},k_{3},k_{4}}\alpha_{l_{1},k_{2},l_{3},l_{4}}\\\nonumber
&&-
\alpha_{k_{1},l_{2},k_{3},l_{4}}\alpha_{l_{1},k_{2},l_{3},k_{4}}+
\alpha_{k_{1},l_{2},l_{3},k_{4}}\alpha_{l_{1},k_{2},k_{3},l_{4}}-
\alpha_{k_{1},l_{2},l_{3},l_{4}}\alpha_{l_{1},k_{2},k_{3},k_{4}}
|^{2}
\\\nonumber&&+
|-\alpha_{k_{1},k_{2},k_{3},k_{4}}\alpha_{l_{1},l_{2},l_{3},l_{4}}+
\alpha_{k_{1},k_{2},k_{3},l_{4}}\alpha_{l_{1},l_{2},l_{3},k_{4}}+
\alpha_{k_{1},k_{2},l_{3},k_{4}}\alpha_{l_{1},l_{2},k_{3},l_{4}}\\\nonumber
&&-
\alpha_{k_{1},k_{2},l_{3},l_{4}}\alpha_{l_{1},l_{2},k_{3},k_{4}}
-\alpha_{k_{1},l_{2},k_{3},k_{4}}\alpha_{l_{1},k_{2},l_{3},l_{4}}+
\alpha_{k_{1},l_{2},k_{3},l_{4}}\alpha_{l_{1},k_{2},l_{3},k_{4}}\\\nonumber
&&+
\alpha_{k_{1},l_{2},l_{3},k_{4}}\alpha_{l_{1},k_{2},k_{3},l_{4}}-
\alpha_{k_{1},l_{2},l_{3},l_{4}}\alpha_{l_{1},k_{2},k_{3},k_{4}}
|^{2}])^{1/2},
\end{eqnarray}
where $\mathcal{N}^{GHZ}_{4}$ is a normalization constant. As an
example let us investigate the concurrence for the $GHZ^{3}$ class
of four-qubit state. Let
$\beta_{1}=\alpha_{1,1,1,1}\alpha_{2,2,2,2}$,
$\beta_{2}=\alpha_{1,1,1,2}\alpha_{2,2,2,1}$,
$\beta_{3}=\alpha_{1,1,2,1}\alpha_{2,2,1,2}$,
$\beta_{4}=\alpha_{1,1,2,2}\alpha_{2,2,1,1}$,
$\beta_{5}=\alpha_{1,2,1,1}\alpha_{2,1,2,2}$,
$\beta_{6}=\alpha_{1,2,1,2}\alpha_{2,1,2,1}$,
$\beta_{7}=\alpha_{1,2,2,1}\alpha_{2,1,1,2}$,
$\beta_{8}=\alpha_{1,2,2,2}\alpha_{2,1,1,1}$, then we have
\begin{eqnarray}\nonumber
  && \mathcal{C}(4\mathcal{Q}^{GHZ^{4}}_{4}(2,\ldots,2)=(4\mathcal{N}^{GHZ}_{4}
 [
|-\beta_{1}- \beta_{2}- \beta_{3}- \beta_{4} +\beta_{5}+
\beta_{6}\\\nonumber &&+ \beta_{7}+ \beta_{8}
|^{2} 
+|-\beta_{1} -\beta_{2}+\beta_{3} +\beta_{4} -\beta_{5}-
\beta_{6}+ \beta_{7}+ \beta_{8}
|^{2}
\\\nonumber&&+
|-\beta_{1}+ \beta_{2}- \beta_{3}+ \beta_{4} -\beta_{5}+
\beta_{6}+ \beta_{7}- \beta_{8}
|^{2}+
|-\beta_{1}- \beta_{2}\\\nonumber &&+ \beta_{3}+ \beta_{4}
+\beta_{5}+ \beta_{6}- \beta_{7}- \beta_{8}
|^{2} 
+|-\beta_{1}+ \beta_{2}- \beta_{3}+ \beta_{4}
+\beta_{5}\\\nonumber &&- \beta_{6}+ \beta_{7}- \beta_{8}
|^{2}
+ |-\beta_{1}+ \beta_{2}+ \beta_{3}- \beta_{4} -\beta_{5}+
\beta_{6}+ \beta_{7}- \beta_{8} |^{2}])^{1/2}.
\end{eqnarray}
Next, let us consider following the state
$\ket{\Psi^{4}_{GHZ}}=\alpha_{1,1,1,1}\ket{1,1,1,1}+\alpha_{2,2,2,2}\ket{2,2,2,2}
$ then the concurrence of the $GHZ^{4}$ class  of the
$\ket{\Psi^{4}_{GHZ}}$ state is given by
\begin{equation}
\mathcal{C}(\mathcal{Q}^{GHZ^{4}}_{4}(2,\ldots,2))=
    (4 \cdot 6\mathcal{N}^{GHZ}_{4}|\alpha_{1,1,1,1}\alpha_{2,2,2,2}
    |^{2})^{1/2}
\end{equation}
 and for
    $\alpha_{1,1,1,1}=\alpha_{2,2,2,2}=\frac{1}{\sqrt{2}}$ we get
    $\mathcal{C}(\mathcal{Q}^{GHZ^{4}}_{4}(2,\ldots,2))=
    (6\mathcal{N}^{GHZ}_{4})^{1/2}$.

\section{Conclusion}
In this paper we have expressed concurrence for a general pure
bipartite state based on an orthogonal complement of our POVM.
Moreover, we have proposed different concurrence classes for pure
multipartite states. We have investigate the monotonicity of the
$W^{m}$ class and the $GHZ^{m}$ class concurrences for multi-qubit
states. The $W^{m}$ class concurrence  for multi-qubit states are
entanglement monotones. However, $GHZ^{m}$ class concurrences are
not entanglement monotones. Our classification suggested  the
existence different classes of multipartite entanglement which are
in equivalent under LOCC. At least, we known that there is two
different classes of entanglement for multi-qubit states which our
methods could distinguish very well. For higher dimensional
composite states, e.g., for $m$-partite states for $m\geq4$, there
is no well known and well accepted classification. Thus, there is
more space for new idea and methods to gives a rigorous
classification of multipartite states. However, we think that this
work is a timely contribution to the relatively large effort
presently being undertaken to quantify and classify multipartite
entanglement.
\begin{flushleft}
\textbf{Acknowledgments:} The author acknowledges useful comments
from Jonas S\"{o}derholm and useful discussions with Gunnar
Bj\"{o}rk. This work was supported by the Wenner-Gren Foundations.
\end{flushleft}


\end{document}